\documentclass[useAMS,usenatbib]{mn2e}
\usepackage{graphicx}
\usepackage{pstricks}
\usepackage{psfrag}
\usepackage{wrapfig}
\usepackage{xspace}
\usepackage{paralist} 

\newcommand{\apj}{    {\it Astrophys. J.}}

\newcommand{\mnras}{  {\it Mon. Not. Roy. Astron. Soc.}}
\newcommand{\nat}{    {\it Nature}}

\newcommand{\solphys}{{\it Solar Phys.}}

\title[Magnetic field variations and the seismicity of solar active regions]{Magnetic field variations and seismicity of solar active regions}
\author[J. C. Mart\'inez-Oliveros and A.-C. Donea]{J. C. Mart\'inez-Oliveros\thanks{E-mail:
 Juan.Oliveros@sci.monash.edu.au} and  A.-C. Donea \\
Centre for Stellar and Planetary Astrophysics, School of Mathematical Sciences, Monash University, Victoria 3800, Australia}

\begin{document}

\date{Accepted 1988 December 15. Received 1988 December 14; in original form 1988 October 11}

\pagerange{\pageref{firstpage}--\pageref{lastpage}} \pubyear{2002}

\maketitle

\label{firstpage}

%
%
%
%

\begin{abstract}

Dynamical changes in the solar corona have proven to be very important in inducing seismic waves into the photosphere. Different mechanisms for their generation have been proposed. In this work, we explore the magnetic field forces as plausible mechanisms  to generate sunquakes as proposed by \citeauthor{hfw2007}. We present a spatial and temporal analysis of the line-of-sight magnetic field variations induced by the seismically active 2003 October 29 and 2005 January 15 solar flares and compare these results with other supporting observations.

\end{abstract}



\section{Introduction}

Over time the study of the Sun has revealed the relationship between the magnetic field inside the Sun and the processes that occur in the upper layers of the solar atmosphere. Apparently, this relationship works in both directions. Recently, the influence of flares on the solar photosphere and the solar interior has been determined \citep{kz1998,dl2005,donea2006,moradi07,Martinez-Oliveros2007,Martinez-Oliveros2008a}. This relationship is partly seen in the seismicity of the  active regions that hosted the flare. In this context, seismicity is defined as the surface manifestation of the refracted back to the surface acoustic waves induced by the flare.

Various scenarios have been proposed to explain the solar seismicity, and in particular to explain the absence of seismic activity in a majority of flares. Under these assumptions, sunquakes can be generated by different physical processes:
\begin{inparaenum}[\itshape i)]
\item  \citet{kz1998} argue that  sunquakes are generated by chromospheric shocks resulting from the explosive ablation of the chromosphere by high-energy electrons. This shock propagates through the photosphere and into the solar interior;
\item \citet{donea2006} suggested that the sudden heating of the low photosphere   (back-radiative warming \citep{machado1989})  contributed to seismic emission into the solar interior;
\item the direct interaction of high-energy protons with the photosphere, observed in some flares seismically active, has also been proposed as a possible mechanism of seismic excitation \citep{dl2005,zharkova2007};
\item finally, \citet{hfw2007}  have recently introduced an alternative idea for the coupling of flare energy into a seismic wave, namely the ``McClymont magnetic jerk'' produced during the impulsive phase of acoustically active flares. The jerk is produced when the coronal loop collapses and the magnetic field lines relax, reducing the total amount of magnetic free energy in the system. This relaxation makes the field lines in the photosphere become ``more horizontal''. Therefore, changes in the Lorentz force on the photosphere may excite acoustic waves into the surface and subsurface of the Sun. Based on this work and previous results from \citet{sh2005} and \citet{h2000}, \citet{hfw2007} estimated the mechanical work applied to the photosphere by a sudden coronal restructuring. The energy estimates are similar to those based on our helioseismic observations \citep{dl2005} and suggest that the ``McClymont Jerk'' can account for the seismic activity of solar flares.
\end{inparaenum}

In this Letter we expand upon the work of \citet{sh2005} on flare-associated magnetic field changes  by  making  temporal and spatial studies  of the line-of-sight magnetic field  of two very well studied active regions associated with sunquakes. We follow the steps described in \citet{sh2005}, using data from the Michelson Doppler Imager (MDI) onboard the Solar and Heliospheric Observatory (SOHO), and compare these results with helioseismic holography images for the specified flares \citep{dl2005,moradi07,Martinez-Oliveros2008b} and Global Oscillations Network Group (GONG++) Intensity continuum.

This Letter adds new results to our previous work \citep{Martinez-Oliveros2008b}. Our aim is to determine whether there are any persistent changes in the magnetic field associated with two major flare generated seismic sources.  

\section{Observations and Analysis}

\begin{figure}
\includegraphics[width=1.\columnwidth]{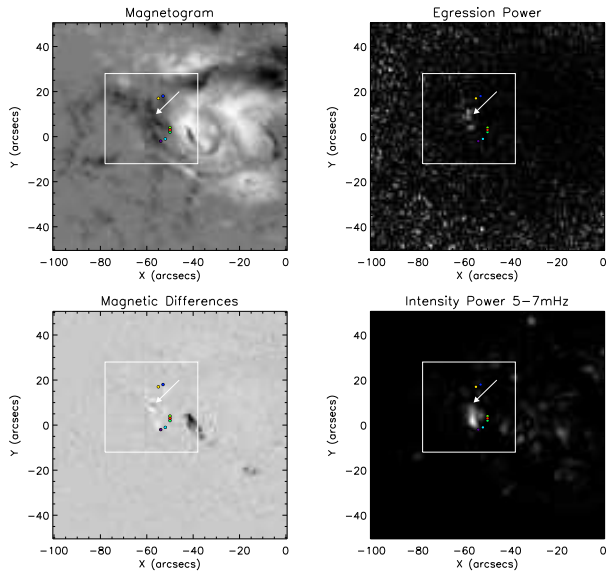} 
\caption{Top Left: SOHO--MDI magnetogram of the active region AR10486 at 20:43~UT. Top Right: 5--7mHz egression power map at 20:46~UT. Bottom Left: SOHO--MDI magnetogram differences at 20:43:30~UT. Bottom Right: 5--7mHz GONG++ Intensity continuum power map at 20:44~UT. The area analysed is represented by the white box. The arrow points to the main acoustic kernel.}
\label{FIG1}
\end{figure}

Sunquakes, are a phenomena associated with highly impulsive and abrupt processes. So, in order to determine the influence of the variation of magnetic field (McClymont Jerk) in their generation it is necessary to correlate the different spatial and temporal properties of different signals (magnetic, acoustic, etc.). First, we created time variation plots of the magnetic field as described in \citet{sh2005}, looking for changes in the line-of sight magnetic field, which could lead to a seismic response of the solar photosphere. We use for this study  SOHO-MDI datasets which were remapped to a postel projection centred at the seismic source. Then, we analysed an area of  $\mathrm 40\times 40$ pixels   covering the main features of the seismic source. We employed maps of acoustic power and egression power of the seismic sources as detailed in \citet{dl2005,moradi07,Martinez-Oliveros2008b}.

\begin{figure}
\includegraphics[width=1.\columnwidth]{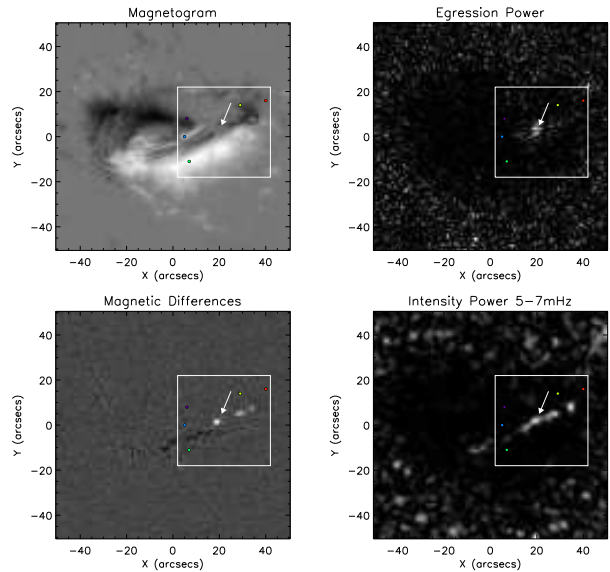} 
\caption{Top Left: SOHO--MDI magnetogram of the active region AR10720 at 00:44~UT. Top Right: 5--7mHz egression power map at 00:48~UT. Bottom Left: SOHO--MDI magnetogram differences at 00:44:30~UT. Bottom Right: 5--7mHz GONG++ Intensity continuum power map at 00:40~UT. The area analysed is represented by the white box. The arrow shows the position of the main acoustic kernel.}
\label{FIG2}
\end{figure}

From previous investigations \citep{kz2001, sh2005}, we know that the time variation of the magnetic field changes during flares can be represented by a first order step-like function. Following \citet{sh2005}, we fitted for each individual pixel in the analysed $\mathrm 40\times 40$ pixels area, the temporal profile  of the magnetic field  with the following step-like function using a Levenberg-Marquardt least-squares minimisation algorithm,

\begin{equation}
\label{eq1}
B(t) = a +bt +c\left\lbrace1+\frac{2}{\pi}\tan^{-1}[n(t-t_0)] \right\rbrace
\end{equation}

\noindent where $a$ and $b$ account for the strength and the evolution of the background field, $t_0$ is step half-time, $c$ is the half-size of the step and $n$ is related to the slope of the step function. As we are interested in  correlations between the seismicity presented by the active region and the change of the magnetic field, we restrict our analysis to those pixels for which the value of the parameter $t_0$ was range between $\pm 10$~minutes from the time reported by GOES as the maximum. 

Then, we have  generated a map with pixel-size areas where we measured step-like magnetic field variations.  Accordingly to \citet{hfw2007} these areas are good candidates for the generation of sunquakes. If these changes are associated with the generation of sunquakes they should be located in regions where the excess of acoustic power is observed. We are going to apply the best-fitting method described in equation~(\ref{eq1}) to two seismic events that occurred on 2003 October 29 and 2005 January 15.

\subsection{The highly seismic events of  2003 October 29 and 2005 January 15}

In 2003 the active region AR10486 hosted two flares with GOES soft X-ray intensities of X17 and X10 on 28 and 29  October, respectively. \citet{dl2005} studied the seismicity of these flares, detecting  two sunquakes induced by these flares. The seismic region produced on 29 October was composed of a single compact source located close to the east end of the active region as identified in egression power maps obtained by the local technique named helioseismic holography \citep{dbl1999}. The flare began at 20:41~UT, reaching maximum at 20:49~UT and finishing at 21:01~UT. The seismic source developed to its maximum around 20:43~UT as reported by \citet{dl2005}.
Figure~\ref{FIG1} shows the morphology of the magnetic AR10486 in the form of a SOHO  MDI magnetogram, an egression power map at 6 mHz and an intensity continuum power map. Magnetic differences during the flare are also shown clearly identifying regions where the magnetic field changed. The area of interest is shown by the white rectangle in Figure~\ref{FIG1}. In all maps one can see the localised response of the photosphere (magnetic, acoustic and white light) to the impulsive flare. Remarkably, there is strong spatial correlation between the seismic source (indicated by the arrow) and the white light power signature \citep{dl2005}.

\begin{figure}
\begin{tabular}{r}
\includegraphics[width=0.98\columnwidth]{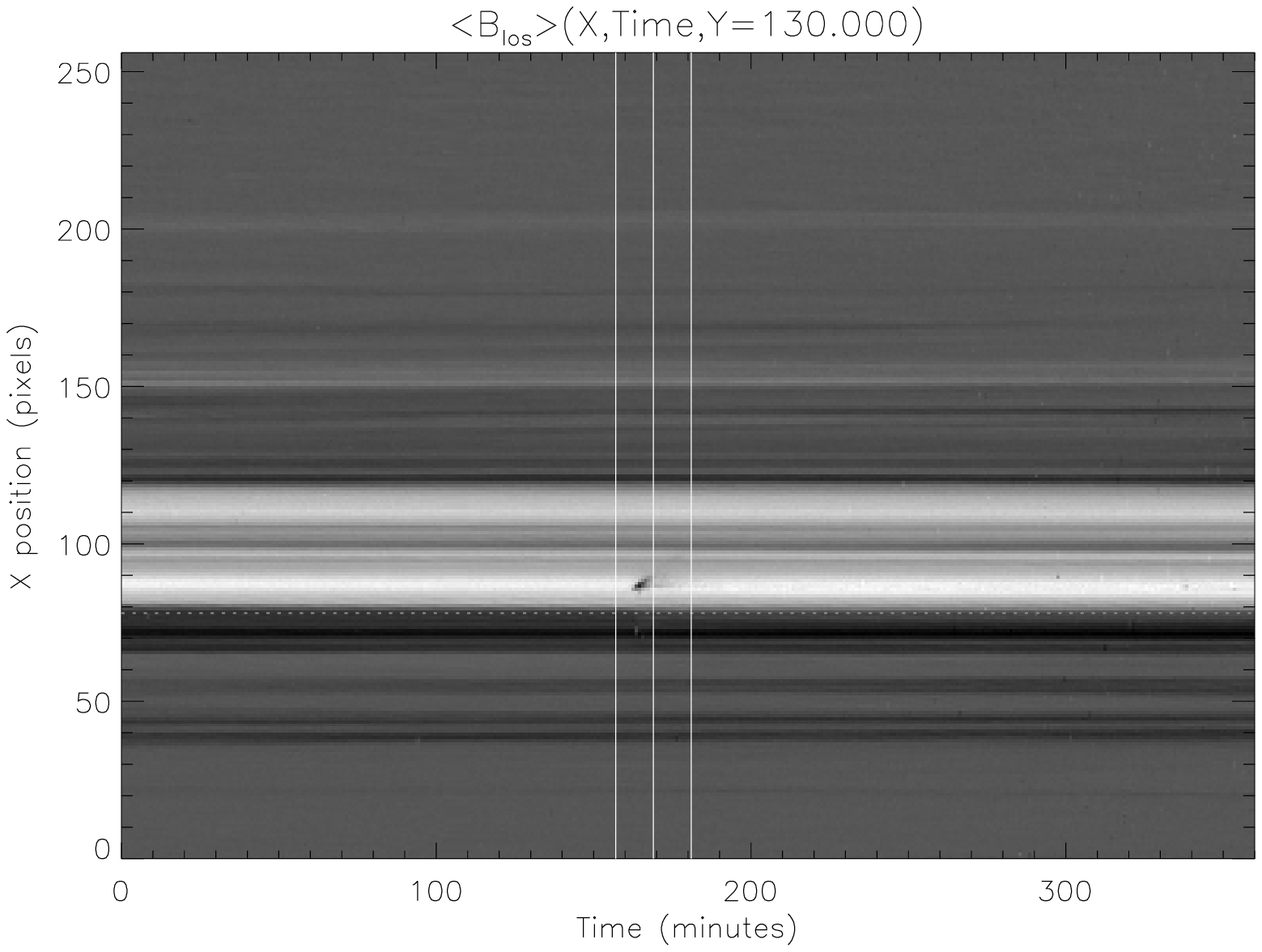} \\
\includegraphics[width=0.98\columnwidth]{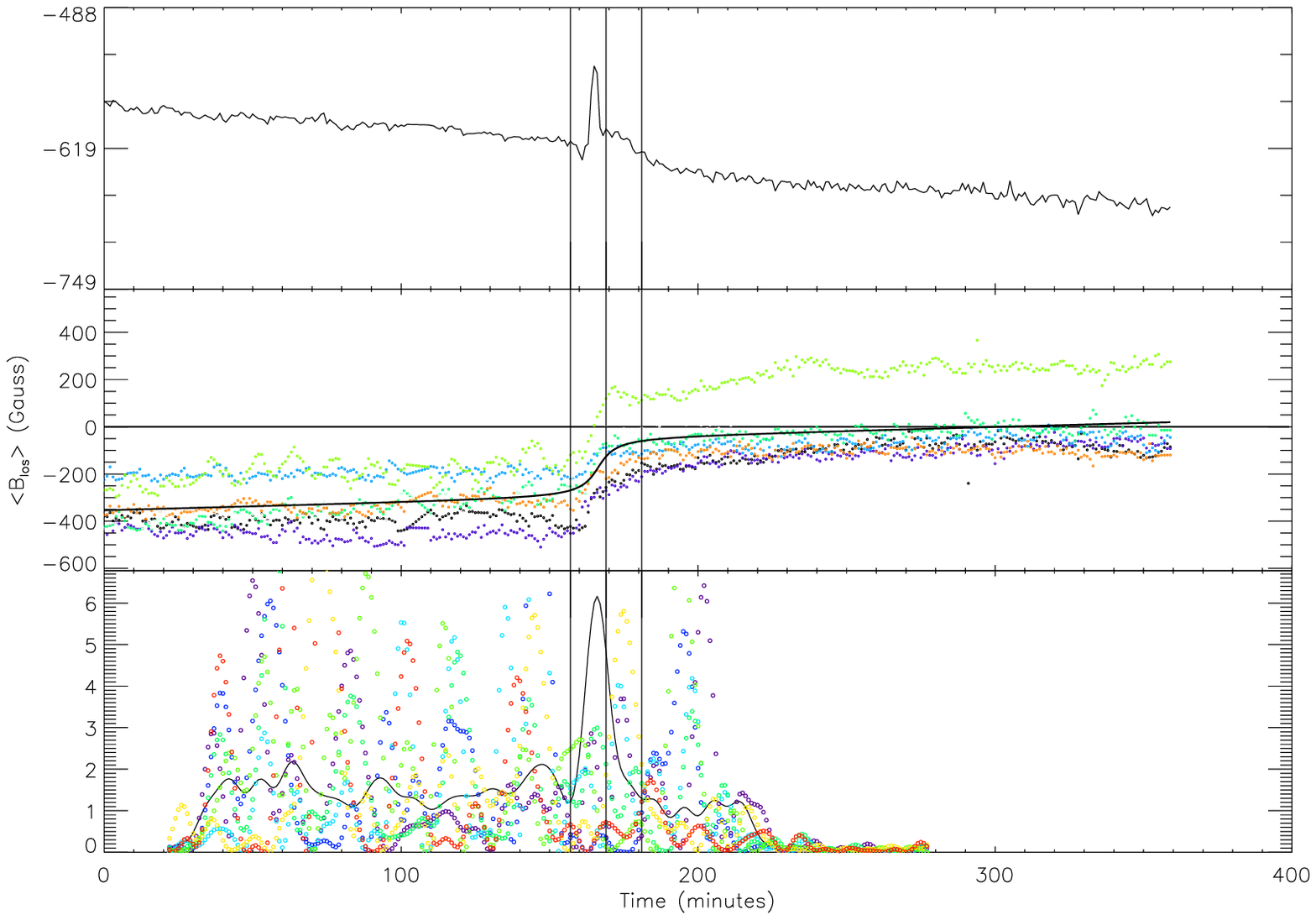} 
\end{tabular}
\caption{\textit{Top image}: l-o-s magnetic field at a fixed y-pixel. The magnetic field transient can be seen as a black spot close to the line representing the time of GOES soft X-ray maximum emission; see below. \textit{Bottom image}: Top frame: SOHO/MDI \textit{l-o-s} magnetic field profile of the main acoustic kernel. The middle frame: Time variation plots of the SOHO/MDI \textit{l-o-s} magnetic field of representative points from figure~\ref{FIG1}. A fit to the mean value of the data is plotted as well. Bottom frame: Time series of the 5--7~mHz normalised egression power of each representative point. The solid curve represents the egression power integrated over the main acoustic kernel. The three vertical lines in both plots represent the start (20:41~UT), maximum (20:49~UT) and end (21:01~UT) of the solar flare according to the GOES X-ray flux.}
\label{FIG3}
\end{figure}

Following the studies of \citet{sh2005} we have examined the temporal profile of the  magnetic field over each pixel in the  area shown by the rectangle in Figure~\ref{FIG2}. Similarly, step-like changes in the magnetic field have been identified in some pixel-size areas of this active region.  One MDI pixel represents about 1.4 Mm spatial resolution. Figure~\ref{FIG1} shows the  relative position of these pixels in maps  of  different  emissions.  We found that  no pixel with a clear step-like  behaviour of the magnetic field appears  inside the detected seismic regions. This results casts doubt on whether  the ``McClymont magnetic jerk'' can account for  the seismic activity of the 2003 October 29 solar flare.

\begin{figure}
\begin{center}
\begin{tabular}{c}
\includegraphics[width=0.98\columnwidth]{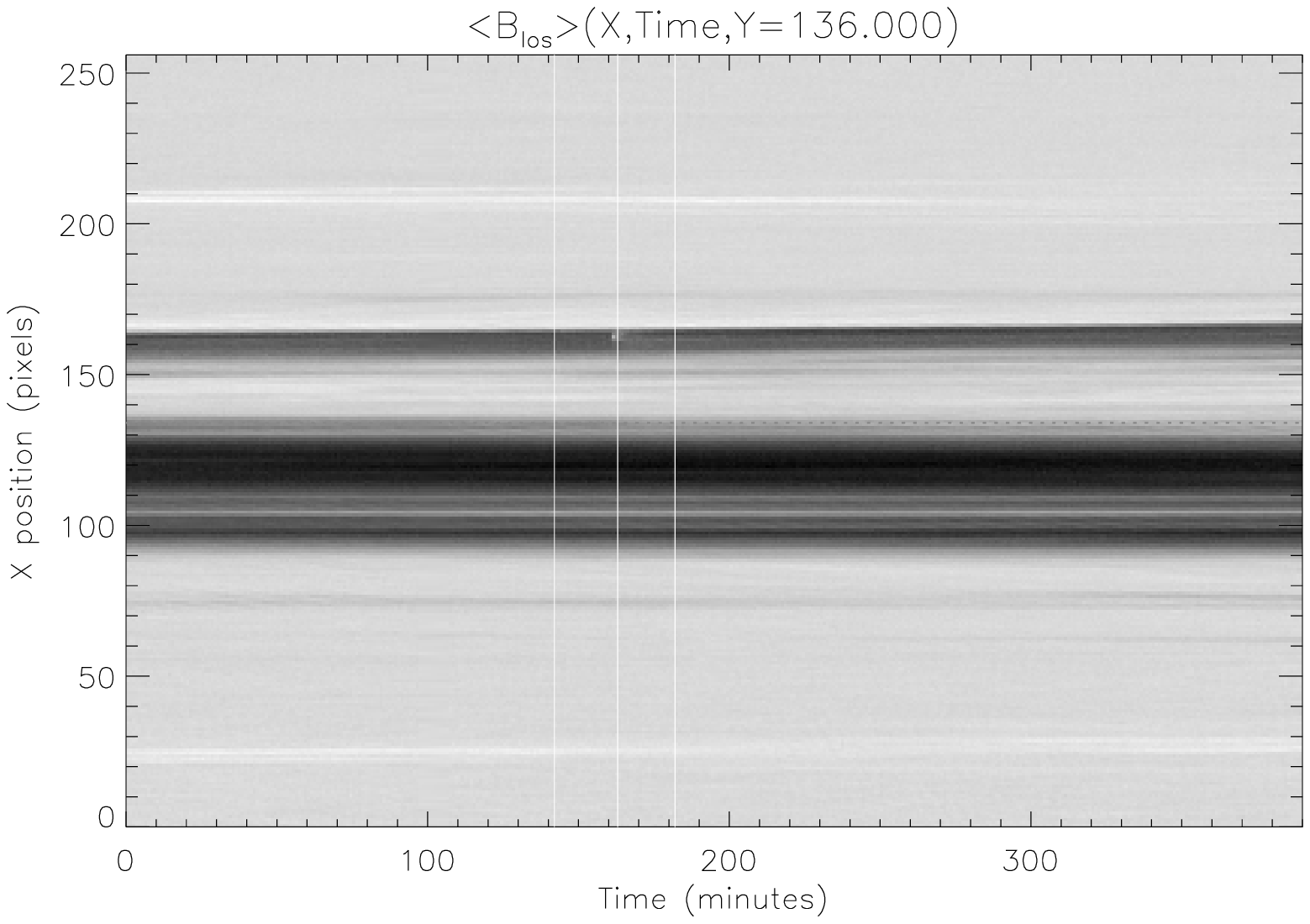} \\
\includegraphics[width=0.98\columnwidth]{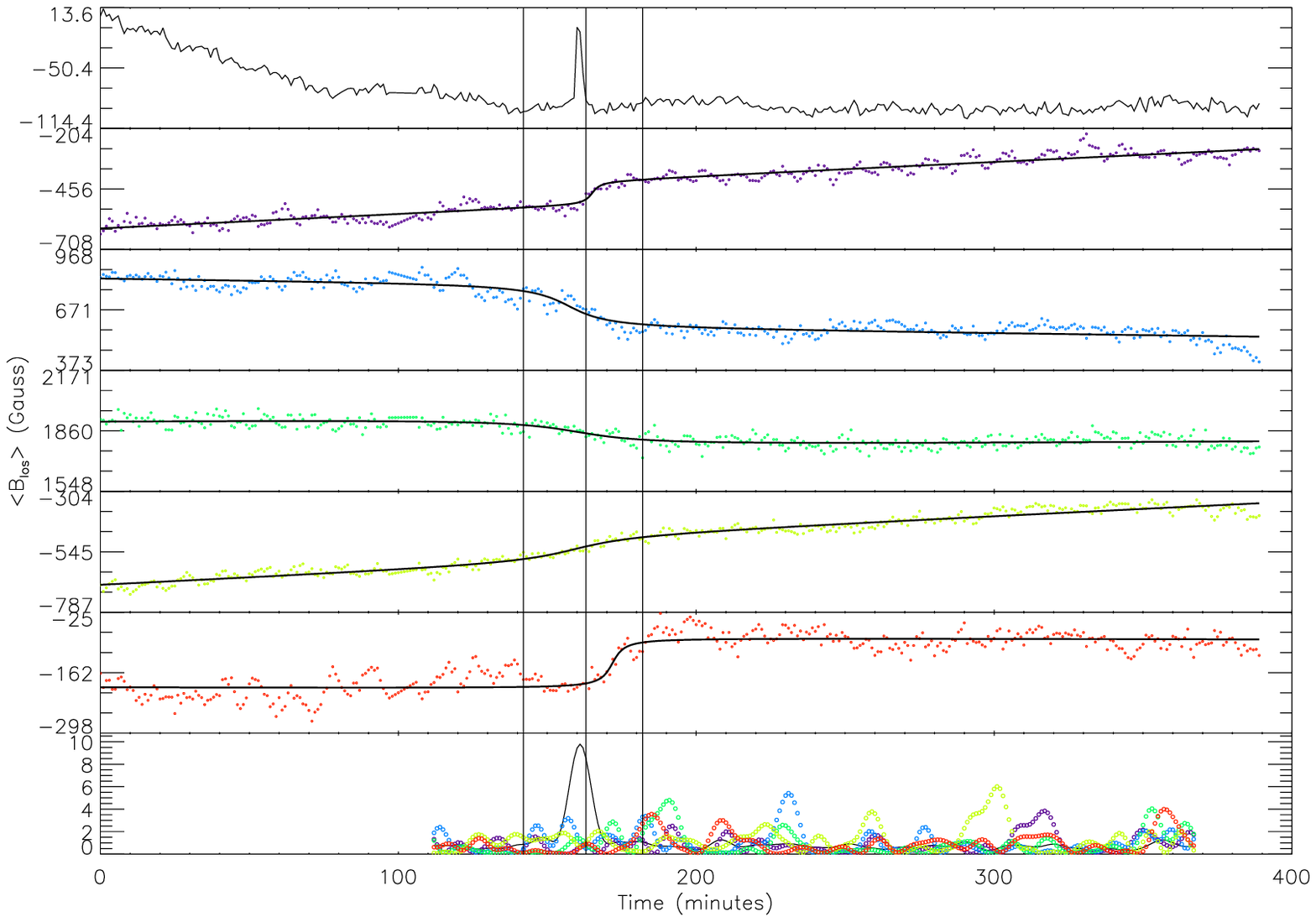} 
\end{tabular}
 \end{center}
\caption{\textit{Top image}: l-o-s magnetic field at a fixed y-pixel. The magnetic field transient can be seen clearly as a white spot under the solid line representing the time of GOES soft X-ray maximum emission, see below. \textit{Bottom image}:
Top frame: SOHO/MDI \textit{l-o-s} magnetic field profile of the main acoustic kernel. The five middle frames: Time variation plots of the SOHO/MDI \textit{l-o-s} magnetic field of representative points as discussed in the text. A fit to the mean value of the data is plotted as well. Bottom frame: Time series of the 5--7~mHz normalised egression power of each representative point. The solid curve represents the egression power integrated over the main acoustic kernel. The three vertical lines in both plots represent the start (00:22~UT), maximum (00:43~UT) and end (01:02~UT) of the solar flare according to the GOES X-ray flux.}
\label{FIG4}
\end{figure}

The second flare of interest, occurred on 2005 January 15 in the  active region AR10720, which  produced  5 more X-class solar flares. \citet{betal2006} and \citet{Moradi2006a,Moradi2006b} reported a powerful seismic transient generated by the X1.2 flare of January 15. Properties of the seismic ripples  generated by this event were later analysed by \citet{kz2006}.  \citet{moradi07}  and \citet{Martinez-Oliveros2008b}  compared the seismic source morphology  with other supporting observations.   They  emphasised the spatial coincidence between the strong compact acoustic source and signatures of hard X-ray emission, suggesting that high-energy electrons played an important role in triggering the seismic event.  Also,  it was shown that the acoustic emission occurred at the location of  conspicuous  white-light signatures, suggesting that the radiative back-warming mechanism was also relevant  in transporting energy into the low photosphere.

Using the best-fitting method described in equation~(\ref{eq1}) we have searched for any abrupt and permanent changes in the magnetic field of the seismic areas of AR10720 shown by the arrow in Figure~\ref{FIG2}.  We have detected few points with a significant  step-like function ($\mathrm  \Delta B \approx 250$ G) offset from the seismic structures.  A fitting parameter $t_0=\mathrm \pm 10$ minutes off the flare maximum (00:43~UT) has been taken. In addition, the steepening happened in a gradual manner and it was  not sudden as required for the generation of seismic waves  \citep{dl2005}.

Temporal profiles of the  line-of-sight (l-o-s) magnetic field  of the representative points and of the highly seismic regions are shown in Figures~\ref{FIG3} and~\ref{FIG4}. We also include the best-fit of equation~(\ref{eq1}) to the plots to illustrate the field changes. Most of the pixels in the maps showing fast changes in the magnetic field (left bottom panels in Figures~\ref{FIG1} and~\ref{FIG2}) display transients. We also show the time variation of the egression power, which is the signature of any seismic source at the corresponding location. As the representative points are located in regions of the outer penumbra or quiet sun, the 6~mHz egression power of each pixel becomes comparable with the emission of the sunquake. This fact is clearly seen in Figure~\ref{FIG3}. 

\section{Discussion}

In the last year, seismicity induced by solar flares of different intensities, from M6.7 to X17, has been observed \citep{dl2005,donea2006,moradi07,kz1998,Martinez-Oliveros2007,Martinez-Oliveros2008a}. Different mechanisms of generation for seismic waves have been proposed and it is clear that in all of them the magnetic field plays not only an important, but a decisive role. The dynamical character of the magnetic field is perhaps a principal phenomenon in all of these mechanisms. These dynamic changes are, according to \citet{hfw2007}, responsible for the seismic emissions observed on the solar surface. 

We analysed two seismically active solar flares, and studied the variations of the line-of-sight magnetic field looking for sudden changes in the value of the magnetic field strength. We studied the temporal and spatial properties of the pixels associated with these variations and their correlation with proxies of the seismicity, such as excess in white light emission and magnetic transients. In both cases we found a good temporal correlation between the magnetic field variations and the maximum of the GOES soft X-rays flux. However, we did not find a good spatial correspondence between the representative points and the sunquakes acoustic emission at 5--7~mHz; and what is more interesting, these pixels do not correlate spatially with the magnetic transients observed in the magnetic differences. In the case of the 2003 October 29 flare, the representative points are localised near the seismic source, similar to what \citet{sh2005} identified. The 2003 October 29 solar flare shows the desired behaviour for the McClymont jerk effect ``to work'', a sudden change in the l-o-s magnetic field strength. In the case of the 2005 January 15 solar flare, the representative points are dispersed all over the  analysed region, presenting no spatial correlation with the seismic sources. These results show that a substantial step-like change in the magnetic field strength cannot always be associated with a  seismic source during a flare. Transient magnetic events seem to dominate the seismic areas of the 2005 January flare. They  have also been  detected in 
a number of flares, some of which were acoustically active and others of which were not (detectably). These magnetic 
signatures are spatially and temporally consistent with the acoustic signatures. However, we question the reliability of the magnetic signatures during and some time following a white light flare.

Another interesting aspect is related to the suddenness effect: Are the impulsive changes of the magnetic field of the order of minutes (10 minutes in the cases of the 2003 October 29 and 2005 January 15 solar flares) sudden enough to generate seismic waves via the Lorentz magnetic force?

 Limited by the existing observations, we conclude that the  ``McClymont jerk'' mechanism may still  contribute, in parallel with 
the chromospheric shocks driven by sudden, thick-target heating of the upper and middle chromosphere \citep{kz1998,dl2005} and the ``back-warming,'' to building enough strength  to trigger a seismic source.

These results also show that it  is necessary a more detailed study of the magnetic field fine structures and their temporal variation is required, making use of the  photospheric vector magnetograms as provided for example by {\it  Hinode}. This will make it possible to study the real changes in the magnetic field configuration and structure. Eventually these kinds of studies could lead us to an understanding of the role of magnetic forces in enhancing the seismicity of flaring active regions.

\vspace{5mm}
\thanks{\textbf{Acknowledgement}: The authors would like to sincerely thank Prof. Paul Cally and Dr. Charlie Lindsey for their helpful
and interesting comments that helped the progress of this work.}

\label{lastpage}

\end{document}